\documentclass[aps,pra,balancelastpage,preprintnumbers,noflushbottom,twocolumn,reprint,showpacs,floatfix]{revtex4-1}

\usepackage{amsmath,amssymb}
\usepackage{graphicx}
\usepackage{bm}

\newcommand{\TR}{\text{Tr}}
\newcommand{\mydagger}{{+}}
\newcommand{\phdagger}{{\phantom{\mydagger}\!\!\!\!}}
\newcommand{\CRE}{c^\mydagger}
\newcommand{\ANN}{c^\phdagger}
\newcommand{\bra}[1]{\langle{#1}|}
\newcommand{\ket}[1]{|{#1}\rangle}

\newcommand{\expval}[1]{\langle{#1}\rangle}
\newcommand{\RE}{\text{Re}}
\newcommand{\IM}{\text{Im}}
\newcommand{\BG}{{\bm{G}}}
\newcommand{\BK}{{\bm{k}}}
\newcommand{\BR}{{\bm{R}}}

\begin{document}

  \title{Kinetic description of thermalization dynamics in weakly
    interacting quantum systems}
  
  \author{Michael Stark}
  \author{Marcus Kollar}
  \affiliation{Theoretical Physics III, Center for Electronic
    Correlations and Magnetism, University of Augsburg, 86135 Augsburg,
    Germany}
  
  \date{August 25, 2013}
  
  \begin{abstract}
    After a sudden disruption, weakly interacting quantum systems
    first relax to a prethermalized state that can be described by
    perturbation theory and a generalized Gibbs ensemble. Using these
    properties of the prethermalized state we perturbatively derive a
    kinetic equation which becomes a quantum Boltzmann equation in the
    scaling limit of vanishing interaction. Applying this to
    interaction quenches in the fermionic Hubbard model we find that
    the momentum distribution relaxes to the thermal prediction of
    statistical mechanics. For not too large interaction, this
    two-stage scenario provides a quantitative understanding of the
    time evolution leading from the initial pure via a metastable
    prethermal to the final thermal state.
  \end{abstract}

  \maketitle

  An important insight in the field of thermodynamics, as developed by
  Carnot, Clausius, Kelvin, and others, is that a many-particle system
  in equilibrium can be characterized by only a few parameters such as
  entropy (or, more practically, temperature), volume, and particle
  number. Statistical mechanics, pioneered by Maxwell, Boltzmann,
  Gibbs, and others, predicts the properties of this equilibrium state
  in terms of a statistical operator $\rho$ representing an ensemble
  of identical systems, in which each accessible microstate is equally
  probable~\cite{Balian1991a}.  This statistical approach is very
  successful in describing a wide variety of many-particle systems,
  from classical gases and liquids to quantum matter, such as
  electrons in solids, liquid helium, blackbody radiation, or neutron
  stars.  However, it is an intriguing question how a thermal state of
  an isolated system can in fact evolve from a single
  quantum-mechanical wave function $\ket{\psi}$, because such a pure
  state with density matrix $\rho$ $=$ $\ket{\psi}\bra{\psi}$ cannot
  become a mixed finite-temperature state under the unitary time
  evolution of the Schr\"odiger equation.  If the measured values of
  at least some observables correspond to those of a thermal state, a
  quantum many-body system is said to \emph{thermalize}. On general
  grounds it is clear that for this to happen one must consider
  sufficiently `ergodic' Hamiltonians, large enough systems, and not
  too specially crafted observables. For example, integrable quantum
  systems usually do not thermalize due to the presence of many
  constants of motion, leading to a more detailed memory of the
  initial state than just through its energy, system size, and
  particle number~\cite{Polkovnikov2011RMP}.  Even in nonintegrable
  systems, thermalization is notoriously difficult to observe
  numerically and was established for only a few bosonic and fermionic
  systems~\cite{Kollath2007,Rigol2008a,%
    *Rigol2010f,%
    Eckstein2009a,*Eckstein2010a,Eckstein11,Zhang2012a}.  In recent
  years, remarkable experiments with trapped ultracold atomic gases
  have become possible, allowing detailed study of well-isolated
  many-particle systems with excellent control over the
  Hamiltonian~\cite{Bloch2008a}, in particular the relaxation of
  integrable~\cite{Kinoshita2006} or weakly nonintegrable
  systems~\cite{Gring2013a}.  Relaxation phenomena of complex quantum
  systems are also observed in time-resolved pump-probe spectroscopy
  on solids~\cite{Zhukov2009,Wall11,Fausti11,Sentef2012pre}.

  \begin{figure}[b]
    \centering
     \includegraphics[height=0.87\hsize,angle=270]{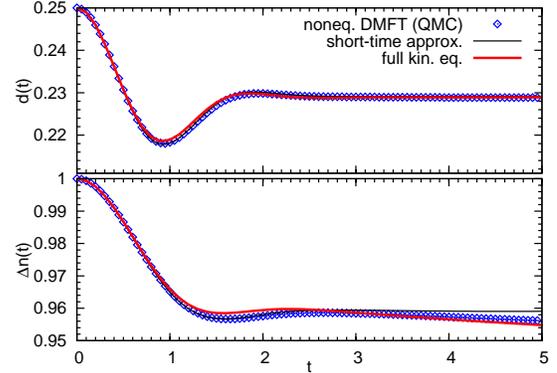}
     \caption{\label{fig:docc+jump}Relaxation dynamics for the Hubbard
       model~\eqref{eq:hubbard} in infinite lattice dimensions with
       semielliptical density of states ($D(\epsilon)$ $=$
       $\sqrt{4-\epsilon^2}/(2\pi)$, bandwidth 4) at half-filling
       after an interaction quench from $U$ $=$ $0$ at time $t$ $=$
       $0$. The other figures also show data for this model. Here the
       quench is to $U$ $=$ $0.5$.  Upper panel: double occupation
       $d(t)$. Lower panel: discontinuity $\Delta n(t)$ of the
       momentum distribution $n_{\bm{k}\sigma}$ $=$
       $n(\epsilon_{\bm{k}\sigma},t)$ at the Fermi surface, $\Delta
       n(t)$ $=$ $n(0^-,t)$ $-$ $n(0^+,t)$.  Symbols: numerically
       exact nonequilibrium DMFT Quantum-Monte-Carlo
       data~\cite{Eckstein2009a}; thin (black) solid lines:
       prethermalization dynamics~\eqref{eq:prethermeq} up to order
       $U^2$~\cite{Moeckel2008a,*Moeckel2009a,*Moeckel2010a,Eckstein2010a}; thick (red) lines:
       full kinetic equation~\eqref{eq:fullkineqint} (with $L$ $=$
       254 discretized momenta).}
  \end{figure}
  A special situation arises for \emph{weakly interacting systems},
  for which the difference between the full Hamiltonian and a
  noninteracting one is small. For such systems a so-called
  prethermalized state during intermediate timescales was predicted by
  perturbative
  methods~\cite{Berges2004a,Moeckel2008a,Kollar2011a,Kitagawa2011a}
  and observed~numerically and
  experimentally~\cite{Eckstein2009a,*Eckstein2010a,%
    Hamerla2013,*Hamerla2013njp,*Hamerla2013b,%
    Gring2013a,%
    Mitra2013a,%
    Marino2013a}.  A crossover from this prethermal to the thermal
  state can then only occur on later timescales.  In this paper we
  develop a theory that describes all timescales, which we evaluate
  and discuss specifically for the thermalization dynamics that occur
  in the fermionic Hubbard model,
  \begin{align}
    \label{eq:hubbard}
    H
    &=
    \sum_{\bm{k}\sigma}
    \epsilon_{\bm{k}}
    \CRE_{\bm{k}\sigma}
    \ANN_{\bm{k}\sigma}
    +
    U
    \sum_{i}
    n_{i\uparrow}
    n_{i\downarrow}
    \,,
  \end{align}
  in the limit of infinite lattice
  dimension~\cite{Metzner1989,Muellerhartmann1989b}. Its equilibrium
  properties can be obtained from dynamical mean-field theory (DMFT)
  which becomes exact in this limit~\cite{Georges96}; in particular,
  this system is a Landau Fermi liquid for sufficiently small $U$,
  with a Mott metal-insulator transition at $U_c$ on the order of the
  bandwidth~\cite{Georges96}. Using nonequilibrium
  DMFT~\cite{Freericks2006}, its time evolution was studied recently
  by numerical~\cite{Eckstein2009a,*Eckstein2010a,Eckstein2010nca,Eurich11,TsujiOkaAokiWerner2012,Gramsch2013pre},
  diagrammatic~\cite{Eckstein10,Eckstein2010nca,Eckstein11,Eckstein2011bloch,TsujiOkaAokiWerner2012,Sentef2012pre,Werner2013wc},
  and variational methods~\cite{Schiro10,*Schiro11}. For intermediate
  interaction quenches the relaxation dynamics changes qualitatively
  at $U_c$ on the order of the
  bandwidth~\cite{Eckstein2009a,*Eckstein2010a,Schiro10,*Schiro11};
  such a dynamical phase transition is observed also in other
  models~\cite{Barmettler2008a,Gambassi2011a,Heyl2012a,Karrasch2013b}.

  Fig.~\ref{fig:docc+jump} shows the time evolution
  of~\eqref{eq:hubbard} after a small quench from $U$ $=$ $0$ to $0.5$
  for a semielliptic density of states (bandwidth~4). While the double
  occupation $d(t)$ $=$
  $\bra{\psi(t)}n_{i\uparrow}n_{i\downarrow}\ket{\psi(t)}$ relaxes
  quickly, the momentum distribution $n_{\bm{k}\sigma}(t)$ $=$
  $\bra{\psi(t)}\CRE_{\bm{k}\sigma}\ANN_{\bm{k}\sigma}\ket{\psi(t)}$
  first prethermalizes and then relaxes further. The relaxation
  towards the prethermalization plateau is given by the perturbative
  result to order $U^2$ (thin solid lines in
  Fig.~\ref{fig:docc+jump}), and the subsequent crossover is described
  by the kinetic theory developed below (thick solid lines).

  \emph{Kinetic equation based on the validity of Wick's theorem.---}
  Quantum kinetic equations were first formulated in the early days of
  quantum mechanics~\cite{Nordheim1928,Ueling1933}, and today the
  literature is
  vast~\cite{Hugenholtz1983a,Erdoes2004a,Bonitzbook1998,*[{}] [{,
      Ch.~3.2.}] Haug2008a,VanVliert08,Spohn08,*Lukkarinen2009a,Snoke12}, including
  recent studies of Luttinger
  liquids~\cite{Lin2013a,Levchenko2013a,Tavora2013a} and the Hubbard
  model~\eqref{eq:hubbard} in one dimension~\cite{Spohn12a,*Spohn13a,*Spohn13b}.
  Mathematically, our approach closely parallels that of
  Ref.~\onlinecite{Erdoes2004a}. In second quantization a generic
  Hamiltonian with two-body interactions reads $H$ $=$ $H_0+gH_1$,
  with
  \begin{align}
    H_0
    &=
    \sum_\alpha\epsilon_\alpha
    \CRE_\alpha\ANN_\alpha
    \,,
    &
    H_1
    &=
    \sum_{\alpha\beta\gamma\delta}
    V_{\alpha\beta\gamma\delta}
    \CRE_\alpha\CRE_\beta\ANN_\gamma\ANN_\delta
    \,,\label{eq:H0H1}
  \end{align}
  where $\CRE_\alpha$ ($\ANN_\alpha$) are creation (annihilation)
  operators, $[\ANN_\alpha,\CRE_\beta]_{-\eta}^\phdagger$ $=$
  $\delta_{\alpha\beta}$, with $\eta$ $=$ $\pm1$ for bosons and
  fermions, respectively; for a hermitian Hamiltonian
  $V_{\alpha\beta\gamma\delta}$ $=$ $\eta V_{\beta\alpha\gamma\delta}$
  $=$ $V_{\delta\gamma\beta\alpha}^*$.
  A natural choice of observables are the occupation numbers
  $n_\alpha$ $=$ $\CRE_\alpha\ANN_\alpha$. We determine the time
  evolution of their expectation value, $N_\nu(t)$ $=$
  $\TR[n_\nu\rho(t)]$. Here the density matrix $\rho(t)$ starts from
  an initial state $\rho(0)$ and obeys the von Neumann equation
  $i\dot{\rho}(t)$ $=$ $[H,\rho(t)]$ (we set $\hbar$ $=$ $1$). By
  iterating this equation we obtain an expression for
  ${dN_\nu(t)}/{dt}$ containing higher-order expectation values
  (see Supplement), which is still exact but requires the (unknown)
  density matrix $\rho(t)$.  We will argue below that we may apply
  Wick's theorem to leading order in $g$ in this equation.  For a
  translationally invariant Hamiltonian and if the initial state is
  the ground state $\ket{\psi_0}$ of $H_0$, i.e., $\rho(0)$ $=$
  $\ket{\psi_0}\bra{\psi_0}$, we then obtain a Volterra integral
  equation of the second kind for $N_\nu(t)$,
  \begin{multline}
    \label{eq:fullkineqint}
    N_\nu(t)
    =
    N_\nu(0)%+g\int\limits_0^th_\nu(s)ds
    -
    16 g^2
    \sum_{\!\beta\gamma\delta\!}
    |V^\phdagger_{\nu\beta\gamma\delta}|^2
    \int\limits_0^t
    \frac{\sin[E_{\nu\beta\gamma\delta}(t-s)]}{E_{\nu\beta\gamma\delta}}
    \\
    \times\,
    (
    N_{\nu}N_{\beta}\bar{N}_{\gamma}\bar{N}_{\delta}
    -
    \bar{N}_{\nu}\bar{N}_{\beta}N_{\gamma}N_{\delta}
    )\Big|_s\,ds
    +O(g^3)
    \,,
  \end{multline}
  with $E_{\nu\beta\gamma\delta}$ $=$ $\epsilon_\nu$ $+$
  $\epsilon_\beta$ $-$ $\epsilon_\gamma$ $-$ $\epsilon_\delta$ and
  $\bar{N}_\alpha$ $=$ $1$ $+$ $\eta N_\alpha$.  This `full kinetic
  equation' can be integrated numerically, see the results below for
  the Hubbard model~\eqref{eq:hubbard}, or used with further
  approximations to obtain the short-time prethermalization or
  long-time Quantum-Boltzmann regimes.

  \emph{Short-time approximation and prethermalization dynamics.---}
  Let us first consider the short-time behavior, i.e., $t$ $\lesssim$
  $\text{const}/g^2$, by expanding $\rho(t)$ = $\rho(0)$ + $O(g)$.
  Then in zeroth order in $g$ Wick's theorem applies in the expression
  for ${dN_\nu(t)}/{dt}$. The occupation numbers
  in~\eqref{eq:fullkineqint} then remain at their initial values at
  $t$ $=$ $0$, so that
  \begin{multline}
    \label{eq:prethermeq}
    N_\nu(t)
    =
    N_\nu(0)
    -
    32 g^2
    \sum_{\!\beta\gamma\delta\!}
    |V^\phdagger_{\nu\beta\gamma\delta}|^2
    \frac{\sin^2(E_{\nu\beta\gamma\delta}t/2)}{(E_{\nu\beta\gamma\delta})^2}
    \\
    \times\,
    (
    N_{\nu}N_{\beta}\bar{N}_{\gamma}\bar{N}_{\delta}
    -
    \bar{N}_{\nu}\bar{N}_{\beta}N_{\gamma}N_{\delta}
    )\Big|_0
    +O(g^3)
    \,.
  \end{multline}
  We remark that for the fermionic Hubbard model ($\eta$ $=$ $-1$), we
  recover from~\eqref{eq:prethermeq} precisely the prethermalization
  dynamics to order $g^2$ that were previously obtained by more
  complicated weak-coupling methods~\cite{Moeckel2008a,Kollar2011a}.
  Note, however, that the validity of the general
  expression~\eqref{eq:prethermeq} is not guaranteed: while it
  predicts the short-time behavior well for the fermionic Hubbard
  model in infinite dimensions, it can lead to unphysical results,
  e.g., in two dimensions~\cite{Hamerla2013b}.
  
  \emph{Beyond the prethermalization regime.---} For the time regime  $t$ $\geq$ const$/g^2$
  we need to keep a time-convoluted  scattering term on the
  right-hand side of the full kinetic equation~\eqref{eq:fullkineqint}.
  We argue for the validity of Wick's theorem, and
  hence~\eqref{eq:fullkineqint}, as follows. The 
  prethermalization plateau, which is given by the the long-time limit
  of~\eqref{eq:prethermeq},  can be obtained by replacing the
  squared sine factor by its average value $\frac12$. The resulting
  metastable value of $N_\nu$ is known to be correctly predicted to order
  $g^2$ by a generalized Gibbs ensemble (GGE)~\cite{Kollar2011a},
  $\rho_{\widetilde{\text{GGE}}}$ $\propto$
  $\exp(\sum_\alpha\lambda_\alpha\tilde{n}_\alpha)$, which is built
  from the approximate constants of motion $\widetilde{n}_\nu$ $=$
  ${n}_\nu$ $+$ $O(g)$ of the weakly interacting Hamiltonian $H_0$ $+$
  $gH_1$.
  In view of the applicability of the GGE we replace the
  exact density matrix $\rho(t)$ by $\rho_{\widetilde{\text{GGE}}}$ at
  the prethermalization timescale, and assume that it approximately
  maintains this form even for longer times, i.e., we set $\rho(t)$
  $\propto$ $\exp(\sum_\alpha\lambda_\alpha(t)\widetilde{n}_\alpha)$.
  This approximate density matrix has the virtue of describing not
  only the prethermalization plateau, but also an expected thermal
  state, for which $\lambda_\alpha$ $=$ $\beta\epsilon_\alpha$.
  Finally we set $\rho(t)$ $\propto$
  $\exp(\sum_\alpha\lambda_\alpha(t)n_\alpha)$ $+$ $O(g)$, which only
  leads to corrections of order $g^3$ in~\eqref{eq:fullkineqint}. This
  argument suggests that $\rho(t)$ is essentially quadratic not only
  up the prethermalization timescale but also beyond, suggesting that
  the use of Wick's theorem in the derivation of~~\eqref{eq:fullkineqint}
  is permissible. Physically this means that for sufficently small $g$, 
  the relaxation is essentially described by the dressed
  degrees of freedom only. Our reasoning might become invalid, e.g., if
  expectation values of the higher-order operators that occur in ${dN_\nu(t)}/{dt}$
  are not described by the approximately quadratic $\rho(t)$.

  \emph{Quantum Boltzmann equation.---} For long times, $t$ $\gg$
  $\text{const}/g^2$, we can introduce the scaled time variable $\tau$
  $=$ $g^2t$ and take the limit $g$ $\to$ $0$~\cite{Hugenholtz1983a,Erdoes2004a}. This
  yields a quantum Boltzmann equation (QBE),
%   \begin{multline}
%     \label{eq:QBE}
%     \frac{dN_\nu(\tau)}{d\tau}
%     =
%     -16\pi
%     \sum_{\!\beta\gamma\delta\!}
%     |V^\phdagger_{\nu\beta\gamma\delta}|^2
%     \,
%     \delta(\epsilon_\nu+\epsilon_\beta-\epsilon_\gamma-\epsilon_\delta)
%     \\
%     \times\,
%     (
%     N_{\nu}N_{\beta}\bar{N}_{\gamma}\bar{N}_{\delta}
%     -
%     \bar{N}_{\nu}\bar{N}_{\beta}N_{\gamma}N_{\delta}
%     )
%     \Big|_\tau
%   \end{multline}
  \begin{multline}
    \label{eq:QBE}
    \!\!\frac{{N_\nu}'(\tau)}{16\pi}
    =
    \sum_{\!\beta\gamma\delta\!}
    |V^\phdagger_{\nu\beta\gamma\delta}|^2
    \,
    \delta(E_{\nu\beta\gamma\delta})
    (
    \bar{N}_{\nu}\bar{N}_{\beta}N_{\gamma}N_{\delta}
    -
    \text{v.v.}%
    )
    \Big|_\tau\!\!\!\!\!\!\!
  \end{multline}
  which is local in the scaled time $\tau$, and thus easy to integrate
  numerically (v.v.\ means interchange of $N$ and $\bar{N}$). For
  sufficiently long times $t$ (and at fixed small $g$), the physical
  occupation numbers $N_\nu(t)$, as obtained from a direct solution
  of~\eqref{eq:fullkineqint}, should tend to $N(\tau=g^2t)$, the
  solution of~\eqref{eq:QBE}.  The limit of $\tau$ $\to$ $0$ on the
  other hand, corresponds to the prethermalization plateau which
  occurs (if at all) at times $t$ on the order of $\text{const}/g^2$.
  Thus we expect that the details of the short-time regime will be
  overlooked by the QBE, which starts at $\tau$ $=$ $0$ from the
  prethermalization plateaus of $N_\nu$. For a thermal Bose or Fermi
  function $n_{th}(\epsilon)$ $=$
  $1/(\exp(\beta(\epsilon_\nu-\mu))-\eta)$ the occupation numbers
  $N_\nu$ become stationary~\cite{Hugenholtz1983a}, which is the
  expected thermalized distribution in the limit of small~$g$.  It is
  straightforward to show that in the QBE regime, the total particle
  number $\sum_{\alpha}N_{\alpha}$ in~\eqref{eq:QBE} (and hence the
  particle number and spin polarization in the fermionic Hubbard
  model) are constant as a function of the scaled time~$\tau$.
  Furthermore the kinetic energy $\sum_\alpha \epsilon_\alpha
  N_\alpha$ is conserved, confirming the prethermalization scenario in
  which kinetic and potential energy but not the momentum distribution
  itself thermalize on short
  timescales~\cite{Berges2004a,Moeckel2008a}. These quantities are
  conserved in~\eqref{eq:QBE} because of the symmetry of the summands
  together with the $\delta$ function in one-particle energies.
  \begin{figure}[t]
    \centering
    \includegraphics[height=\hsize,angle=270,clip]{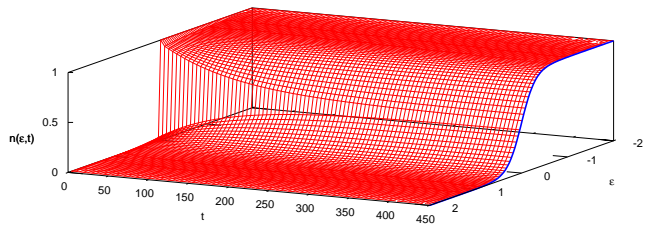}
    \caption{\label{fig:U075netz}%
      Time-dependent momentum distribution
      $n(\epsilon_{\bm{k}\sigma},t)$, as obtained from the
      weak-coupling quantum kinetic equation~\eqref{eq:fullkineqint},
      for a quench from $U$ $=$ $0$ to $0.75$ (with $L$ $=$ 254
      discretized momenta; only a subset of the calculated data points
      are plotted for better visibility). The prethermalization
      plateau is located in the first few grid points at approximately
      $t\lesssim5$.  For large times the momentum distribution
      thermalizes, i.e., it approaches a Fermi function (blue line) at
      the expected temperature $k_BT=0.15$, which corresponds to an
      equilibrium state with the same internal energy as the
      time-evolving system.}
  \end{figure}

\begin{figure}[t]
    \centering
     \includegraphics[height=0.87\hsize,angle=270]{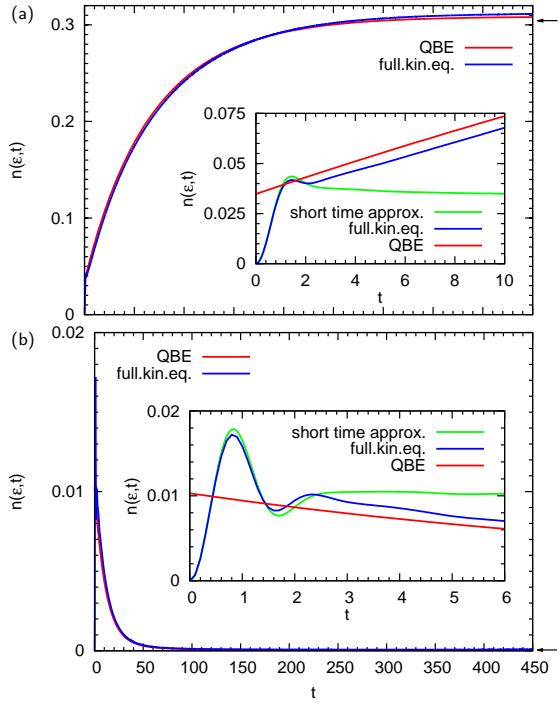}
     \caption{\label{fig:U075eps}%
       Time-dependent momentum distribution $n(\epsilon,t)$ for a
       quench from $U$ $=$ $0$ to $0.75$; (a) $\epsilon=0.1220$ and
       (b) $\epsilon=1.486$. The long-time behavior of the full
       kinetic equation~\eqref{eq:fullkineqint} (with $L$ $=$ 254
       discretized momenta) agrees well with the QBE~\eqref{eq:QBE}
       ($L$ $=$ 1000). The thermal value is
       reached by both at (a) $t\approx 450$ and (b) $t\approx 200$
       (see arrow). Insets: Prethermalization dynamics according to
       the short-time approximation~\eqref{eq:prethermeq} (light green
       line) and full kinetic equation~\eqref{eq:fullkineqint} (blue
       line). The latter only shortly touches the prethermalization
       plateau and then crosses over to the thermal value.}
  \end{figure}

\begin{figure}[t]
    \centering
     \includegraphics[height=0.87\hsize,angle=270]{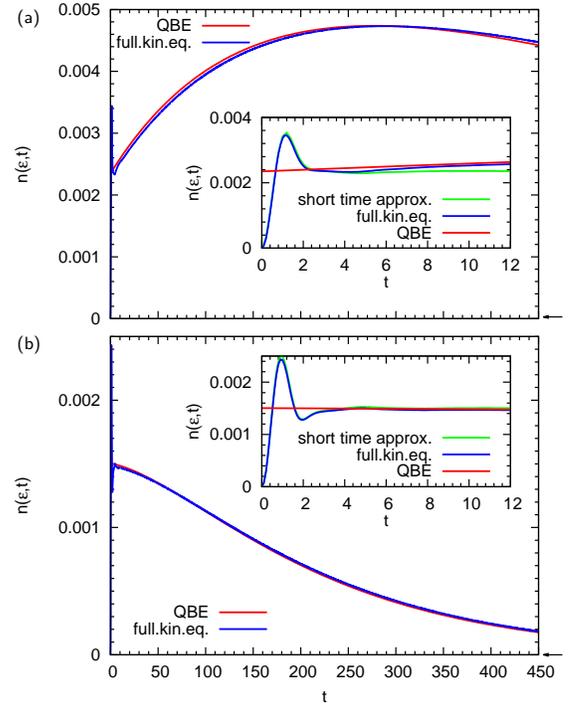}
     \caption{\label{fig:U025eps}%
       As Fig.~\ref{fig:U075eps}, but for quench from $U$ $=$ $0$ to
       $0.25$; (a) $ \epsilon=0.5260$, (b) $\epsilon=1.066$, again
       with good agreement of full kinetic equation and QBE.  Thermal
       value (arrows) is reached by the QBE for (a) $t\approx 10^4$
       and (b) $t\approx 1500$. The crossover to the thermal value is
       slower than in Fig.~\ref{fig:U075eps} because the timescales of
       order $1/U^2$ and $1/U^4$ are better separated.}
  \end{figure}

  \emph{Weak interaction quench in the Hubbard model in infinite
    dimensions.---} We now consider an interaction  quench for the
  fermionic Hubbard model~\eqref{eq:hubbard}. In infinite lattice
  dimensions the evaluation of~\eqref{eq:fullkineqint}-\eqref{eq:QBE}
  simplifies due to the absence of momentum conservation at the
  interaction vertices~\cite{Muellerhartmann1989b}.  In
  Fig~\ref{fig:docc+jump}, the crossover away from the
  prethermalization plateau sets in at $t$ $\gtrsim$ $4$ in the Fermi
  surface discontinuity $\Delta n(t)$ for a quench to $U$ $=$ 0.5.
  Both $\Delta n(t)$ and the double occupation $d(t)$ are described by
  the full kinetic equation \eqref{eq:fullkineqint} in good agreement
  with numerically exact QMC data from nonequilibrium DMFT.  The
  momentum distribution for a quench to $U$ $=$ $0.75$ is shown in
  Fig.~\ref{fig:U075netz}. The initial state $\rho(0)$ is a Fermi sea,
  corresponding to a step function at $t$ $=$ $0$.  After a short
  prethermalization plateau a slow relaxation towards the thermal
  state sets in. At times $t$ $\gtrsim$ $450$ the momentum
  distribution has relaxed to the thermal Fermi function at
  temperature $T$ $=$ 0.15 to high accuracy.  Figs.~\ref{fig:U075eps}
  and Fig.~\ref{fig:U025eps}
  show this behavior in detail for quenches to $U$ $=$ 0.75 and $0.25$
  for two selected band energies $\epsilon$ each.  We observe that the
  data obtained from the full kinetic equation~\eqref{eq:fullkineqint}
  matches both the prethermalization dynamics on short
  timescales~\eqref{eq:prethermeq} and the long-time QBE
  dynamics~\eqref{eq:QBE} in their respective regimes.  This scenario
  for thermalization may also occur in other weakly interacting
  systems, i.e., a full kinetic equation~\eqref{eq:fullkineqint} with
  a prethermalization and a QBE regime and describing the crossover
  between them at intermediate timescales.

\begin{figure}[!b]
    \centering
    \includegraphics[height=0.87\hsize,angle=270]{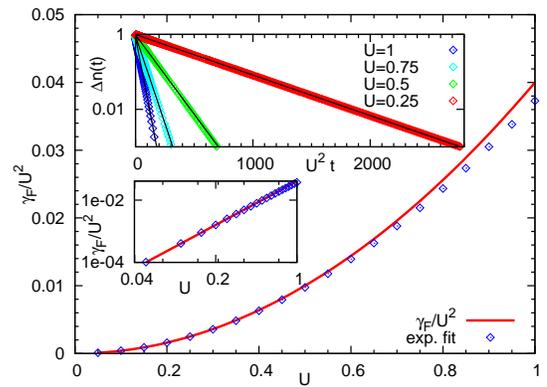}
    \caption{\label{fig:gamma} Relaxation rate $\gamma_{\text{F}}$ at
      the Fermi surface. Upper inset: exponential fit (black lines) to
      QBE data~\eqref{eq:QBE} (symbols) for the Fermi surface
      discontinuity $\Delta n(t)$.  Main panel: quadratic fit in $U$
      for $\gamma_{\text{F}}/U^2$ as extracted from the upper
      inset. Lower inset: logarithmic plot of main panel.}
  \end{figure}

  \emph{Effective temperature and rate of thermalization.---} For the
  fermionic Hubbard model in infinite lattice dimensions, we can
  determine the effective temperature of the final thermal state as
  follows.  As discussed below~\eqref{eq:QBE}, the kinetic energy
  $\expval{H_0}$ is conserved in the QBE regime; we can thus equate
  its expectation value in the prethermalized state (at $\tau$ $=$
  $0^+$) with that in the thermal state (at $\tau$ $=$ $\infty$). A
  Sommerfeld expansion yields $T$ $=$ $aU$, with $a$ determined by the
  density of states (see Supplement).  This agress accurately with the
  relaxed state, e.g., for the quench to $U$ $=$ $0.75$ ($a$ $=$
  $0.20$, $T$ $=$ $0.15$, see Fig.~\ref{fig:U075netz}).
  Furthermore, we can determine the dependence of the relaxation rate
  on $U$ in the QBE regime. For a Fermi liquid, the quasiparticle
  excitations close to the Fermi surface have the longest lifetime and
  therefore relax last.  To determine the asymptotics of
  $n(\epsilon,t)$ for $\epsilon$ $\to$ $0^\pm$ we thus replace the
  distribution function in the integrand of the QBE~\eqref{eq:QBE} by
  thermal Fermi functions $n_{\text{F}}(\epsilon)$ $=$
  $1/(1+e^{\beta\epsilon})$, yielding a rate equation for the distance
  from the thermal state
 ${\delta n^\pm}/{dt}$
    $=$ $-\gamma_{\text{F}}\,\delta n^\pm$ $+$ $O((\delta n^\pm)^2)$
 (after scaling back to $t$ $=$ $\tau/g^2$), where
  $\delta n^\pm(t)$ $=$ $n(0^\pm,t)$ $-$ $\frac12$.
  % \begin{align}
  %   \frac{\delta n^\pm}{dt}
  %   &=-\gamma_{\text{F}}\,\delta n^\pm+O((\delta n^\pm)^2)
  %   \,,
  %   \label{eq:atfermi}
  % \end{align}
  Here $\gamma_{\text{F}}$ $=$ $\pi^3D(0)^3U^2T^2$ in infinite lattice
  dimensions (see Supplement).  QBE data from~\eqref{eq:QBE} for the
  Fermi surface discontinuity $\Delta n(t)$ $=$ $\delta n^-$ $-$
  $\delta n^+$ is shown in Fig.~\ref{fig:gamma}.
  It fits well to an exponential, $\Delta n(t)$ $\propto$
  $\exp(-\gamma_{\text{F}}t)$, with $\gamma_{\text{F}}$ $=$
  $0.038U^4$, in good agreement with the prefactor $a^2$ $=$ 0.040
  predicted from the rate equation for $\delta n^\pm(t)$.  In this
  long-time regime, the system has almost relaxed to thermal
  equilibrium, so that its properties should also be described by
  equilibrium many-body theory. For the particle-hole symmetric
  Hubbard model~\eqref{eq:hubbard} the quasisparticle lifetime
  $\tau_{\bm{k}}$ is obtained from the equilibrium Green function near
  the Fermi surface, $G_{\bm{k}}(\omega)$ $=$
  $1/[\omega-\epsilon_{\bm{k}}-\Sigma_{\bm{k}}(\omega)]$ $\approx$
  $Z_{\bm{k}}/(\omega-Z_{\bm{k}}\epsilon_{\bm{k}}+i/\tau_{\bm{k}})$,
  corresponding to the probability $|G_{\bm{k}}(t)|^2$ $\propto$
  $\exp(-2t/\tau_{\bm{k}})$ of creating a particle or hole in the
  thermal state and removing it after a time $t$. Evaluating
  $\tau_{\text{F}}$ for energies $\epsilon_{\bm{k}}$ at the Fermi
  surface (in infinite lattice dimensions, see Supplement), we obtain
  precisely $\tau_{\text{F}}$ $=$ $2/\gamma_{\text{F}}$ to leading
  order $T^2U^2$; hence the asymptotic timescales for the equilibrium
  and asymptotic nonequilibrium dynamics agree exactly.

  \emph{Conclusion.---} In summary, we showed shown that two known
  weak-coupling paradigms, the prethermalization and the quantum
  Boltzmann regime, are contained in a weak-coupling kinetic theory
  which in particular describes the relaxation in high-dimensional
  fermionic Hubbard models after a small interaction quench well. An
  interesting open question is whether our approach can also describe
  lower-dimensional
  systems~\cite{Hamerla2013,*Hamerla2013njp,*Hamerla2013b}, which
  exhibit qualitatively different short-time dynamics. An application
  to bosonic systems would also be of interest.

   \emph{Acknowledgments.---}
   We would like to thank M.\ %
   Eckstein, S.\ %
   Hamerla, M.\ %
   Heyl, S.\ %
   Kehrein, M. M\"ockel, V.\ %
   Oganesyan, A.\ %
   Polkovnikov, D.\ %
   Snoke, G.\ %
   Uhrig, D.\ %
   Vollhardt, and P.\ %
   Werner for useful discussions. In particular we are grateful to
   Markus Heyl and G\"otz Uhrig for suggestions regarding the
   evaluation of the relaxation rate and its relation to equilibrium
   properties.  This work was supported in part by Transregio~80 of
   the Deutsche Forschungsgemeinschaft.

  % \nocite{*}
  %\bibliographystyle{aipnum4-1}
  

  \appendix

  \onecolumngrid

  \bigskip
  \bigskip

  %\cleardoublepage

  \section*{Supplementary Material}

  \twocolumngrid

  \subsection*{Derivation of the kinetic equations}

  Here we provide details of the derivation of the kinetic equations
  in the main text. For a Hamiltonian of the form~\eqref{eq:H0H1} we
  have
  \begin{align}
    \frac{dN_\nu(t)}{dt}
    &=
    -i\,\TR\,n_\nu[H,\rho(t)]
    =
    \TR\,(-ig)[n_\nu,H_1]\,\rho(t)
    \,.\label{eq:dNdt}
  \end{align}
  In order to iterate the equation of motion we use the interaction
  representation, $\tilde{\rho}(t)$ $=$ $e^{iH_0t}\rho(t) e^{iH_0t}$,
  hence $i\dot{\tilde{\rho}}(t)$ $=$
  $g[\tilde{H}_1(t),\tilde{\rho}(t)]$ and $\tilde{\rho}(0)$ $=$
  $\rho(0)$ $\equiv$ $\rho_0$. After integration this becomes
  $\tilde{\rho}(t)$ $=$ $\rho(0)$ $-ig\int_0^tds
  [\tilde{H}_1(s),\tilde{\rho}(s)]$, so that the expectation
  value of an observable $A$ can be written as
  % \begin{align}
  %   \label{eq:2}
  $\expval{A}_t$
  % &=
  % \TR A\rho(t)
  $=$
  $\expval{\tilde{A}(t)}_0$
  $-$ $ig\int%\limits
  _0^t
  \expval{[\tilde{A}(t\,-\,s),H_1]}_{s}ds$.
  % \,.
  % \end{align}
  Using this with $A_\nu$ $=$ $-ig[n_\nu,H_1]$, setting $g\,h_\nu(t)$ $=$
  $\expval{\tilde{A}_\nu(t)}_0 $ and $g^2I_\nu(t)$ $=$
  $-ig[\tilde{A}_\nu(t),H_1]$, we obtain
  \begin{align}
    \label{eq:ABeq}
    \frac{dN_\nu(t)}{dt}
    &=
    g\,h_\nu(t)    
    -
    g^2\int\limits_0^t\expval{I_\nu(t-s)}_s
    \,
    ds
    \,.
  \end{align}
  Evaluation of these expressions for the Hamiltonian~\eqref{eq:H0H1}
  yields
  \begin{align}
    \label{eq:hfunc}
    h_\nu(t)
    &=
    \IM
    \sum_{\alpha\beta\gamma\delta}
    W_{\alpha\beta\gamma\delta}^{(\nu)}(t)
    \expval{\CRE_\alpha\CRE_\beta\ANN_\gamma\ANN_\delta}_0
    \,,
    \\
    \label{eq:Ifunc}
    I_\nu(t)
    &=
    \sum_{\substack{
        \!\!
        \alpha\alpha'\beta
        \!\!
        \\
        \!\!
        \gamma\gamma'\delta\delta'
        \!\!}
      }
    2W_{\alpha\gamma'\gamma\delta}^{(\nu)}(t)
    V_{\alpha\beta\gamma'\delta'}
    \CRE_\alpha\CRE_\beta[\CRE_{\alpha'},\ANN_{\gamma'}]_\eta^\phdagger\ANN_\gamma\ANN_\delta
    \,,
    % \\
    % W_{\alpha\beta\gamma\delta}^{(\nu)}(t)
    % &=
    % V_{\alpha\beta\gamma\delta}   
    % (\delta_{\alpha\nu}+\delta_{\beta\nu}-\delta_{\gamma\nu}-\delta_{\delta\nu})
    % e^{i(\epsilon_\alpha+\epsilon_\beta-\epsilon_\gamma-\epsilon_\delta)t}
    %\,,
    %\\
  \end{align}%
  with the abbreviations
  \begin{align}%
    W_{\alpha\beta\gamma\delta}^{(\nu)}(t)
    &=
    V_{\alpha\beta\gamma\delta}
    R_{\alpha\beta\gamma\delta}^{(\nu)}(t)
    \,,
    \\
    R_{\alpha\beta\gamma\delta}^{(\nu)}(t)
    &=
    e^{i(\epsilon_\alpha+\epsilon_\beta-\epsilon_\gamma-\epsilon_\delta)t}
    (\delta_{\alpha\nu}+\delta_{\beta\nu}-\delta_{\gamma\nu}-\delta_{\delta\nu})
    \,.
  \end{align}
%   with
%   $R_{\alpha\beta\gamma\delta}^{(\nu)}(t)$
%   $=$
%   $e^{i(\epsilon_\alpha+\epsilon_\beta-\epsilon_\gamma-\epsilon_\delta)t}$
%   $(\delta_{\alpha\nu}$ $+$ $\delta_{\beta\nu}$ $-$ $\delta_{\gamma\nu}$ $-$ $\delta_{\delta\nu})$,
%   $W_{\alpha\beta\gamma\delta}^{(\nu)}(t)$
%   $=$
%   $V^\phdagger_{\alpha\beta\gamma\delta}R_{\alpha\beta\gamma\delta}^{(\nu)}(t)$.
  Note that $h_\nu(t)$ involves only initial state and single-particle
  spectrum $\epsilon_\alpha$; it is absent in particular if $\rho(0)$
  commutes with $n_\nu$, e.g., if the initial state is an eigenstate
  of $H_0$ (this special case is discussed in the main text). Also,
  the inner \mbox{(anti-)}commutator in~\eqref{eq:Ifunc} can be
  written as
  $2\CRE_{\alpha'}\ANN_{\gamma'}+\eta\delta_{\alpha'\gamma'}$, leading
  to one 6-operator and one 4-operator term.  While~\eqref{eq:ABeq} is
  still exact, the integrand involves the (unknown) density matrix
  $\rho(t)$.

  Assuming the validity of Wick's theorem to leading order in $g$ for
  the expectation value of~\eqref{eq:Ifunc}, we obtain
  \begin{align}
    \label{eq:Izero}
    \expval{I_\nu(u)}_s
    &=
    8\sum_{\!\!\alpha\beta\gamma\delta\!\!}
    |V^\phdagger_{\alpha\beta\gamma\delta}|^2
    (\eta+2N_\alpha)N_{\gamma}N_{\delta}
    \RE[R_{\alpha\beta\gamma\delta}^{(\nu)}(u)]
    \nonumber\\&
    ~~~-
    16\eta\sum_{\!\!\alpha\beta\gamma\delta\!\!}
    \RE[V^\phdagger_{\gamma\delta\beta\gamma}
    W_{\alpha\beta\alpha\delta}^{(\nu)}(u)]
    N_{\alpha}N_{\gamma}N_{\delta}
    \,,
  \end{align}
  where the expectation value on the l.h.s. and the densities on the
  r.h.s. are taken at time $s$.  From now on we assume a
  translationally invariant interaction Hamiltonian, for which the
  second line in~\eqref{eq:Izero} vanishes because momentum
  conservation in $V^\phdagger_{\gamma\delta\beta\gamma}$ implies that
  $\beta$ and $\delta$ correspond to the same momentum and hence the
  $W$ factor vanishes~\cite{Erdoes2004a}.  Then~\eqref{eq:ABeq} can be
  written as
  \begin{multline}
    \label{eq:fullkineq}
    \frac{dN_\nu(t)}{dt}
    =
    g\,h_\nu(t)    
    \\
    -
    16 g^2
    \sum_{\!\beta\gamma\delta\!}
    |V^\phdagger_{\nu\beta\gamma\delta}|^2
    \int\limits_0^t
    \,
    \cos[(\epsilon_\nu+\epsilon_\beta-\epsilon_\gamma-\epsilon_\delta)(t-s)]
    \\
    \times\,
    (
    N_{\nu}N_{\beta}\bar{N}_{\gamma}\bar{N}_{\delta}
    -
    \bar{N}_{\nu}\bar{N}_{\beta}N_{\gamma}N_{\delta}
    )\Big|_s
    \,
    ds
    \,,
%     \\
%     [
%     N_{\nu}(s)N_{\beta}(s)\bar{N}_{\gamma}(s)\bar{N}_{\delta}(s)
%     -
%     \bar{N}_{\nu}(s)\bar{N}_{\beta}(s)N_{\gamma}(s)N_{\delta}(s)
%     ]
%     \,,
  \end{multline}
  with $\bar{N}_\alpha$ $=$ $1$ $+$ $\eta N_\alpha$.  For spinless
  fermions ($\eta$ $=$ -1) and $h_v(t)$ $=$ $0$ this kinetic equation
  is the same as the kinetic equation derived in
  Ref.~\onlinecite{Erdoes2004a}, however there the derivation assumed
  $\rho(t)$ to be `quasifree' whereas we assumed the validity of
  Wick's theorem.  In the main text, we argue for the latter on
  physical grounds.  Integrating~\eqref{eq:fullkineq} we obtain a
  Volterra integral equation of the second kind,
  \begin{multline}
    \label{eq:fullkineqint2}
    N_\nu(t)
    =
    N_\nu(0)+g\int\limits_0^th_\nu(s)ds
    \\
    -
    16 g^2
    \sum_{\!\beta\gamma\delta\!}
    |V^\phdagger_{\nu\beta\gamma\delta}|^2
    \int\limits_0^t
    \frac{\sin[(\epsilon_\nu+\epsilon_\beta-\epsilon_\gamma-\epsilon_\delta)(t-s)]}{\epsilon_\nu+\epsilon_\beta-\epsilon_\gamma-\epsilon_\delta}
    \\
    \times\,
    (
    N_{\nu}N_{\beta}\bar{N}_{\gamma}\bar{N}_{\delta}
    -
    \bar{N}_{\nu}\bar{N}_{\beta}N_{\gamma}N_{\delta}
    )\Big|_s
    \,,
%     \\
%     [
%     N_{\nu}(s)N_{\beta}(s)\bar{N}_{\gamma}(s)\bar{N}_{\delta}(s)
%     -
%     \bar{N}_{\nu}(s)\bar{N}_{\beta}(s)N_{\gamma}(s)N_{\delta}(s)
%     ]
%     \,,
  \end{multline}
  which for $h_v(t)$ $=$ 0 reduces to~\eqref{eq:fullkineqint} in the
  main text.

  \subsection{Application to the fermionic Hubbard model}

  For the Hubbard interaction $U\sum_in_{i\uparrow}n_{i\downarrow}$ for
  fermions as in~\eqref{eq:hubbard} on a Brillouin lattice with $L$ sites and periodic
  boundary conditions we set $\alpha$ $=$ $(\BK_1,\sigma_1)$ etc.,
  and $gV_{\alpha\beta\gamma\delta}$ $=$ $\frac{U}{4L}
  \Delta(\BK_1+\BK_2+\BK_3+\BK_4) \sum_\sigma
  \delta_{\sigma_1\sigma} \delta_{\sigma_2\bar{\sigma}} (
  \delta_{\sigma_3\sigma} \delta_{\sigma_4\bar{\sigma}} -
  \delta_{\sigma_3\bar{\sigma}} \delta_{\sigma_4\sigma} )$, where
  $\Delta(\BK)$ $=$ $\sum_\BG$ $\delta_{\BK,\BG}$ $=$ $\frac1L\sum_\BR$ $\exp(i\BK\BR)$ is the von-Laue
  function involving reciprocal lattice vectors $\BG$.%

  For this model the full kinetic equation~\eqref{eq:fullkineqint}, 
  short-time dynamics~\eqref{eq:prethermeq}, and QBE~\eqref{eq:QBE}
  become
  \begin{multline}
    \label{eq:fullkineqinthubb}
    N_{\BK\sigma}
    =
    N_{\BK\sigma}(0)
    -
    \frac{4U^2}{L^2}
    \sum_{\BK_2\BK_3\BK_4}\Delta(\BK+\BK_2-\BK_3-\BK_4)
    \\\times
    \int\limits_0^t
    \frac{\sin[(\epsilon_{\BK}+\epsilon_{\BK_2}-\epsilon_{\BK_3}-\epsilon_{\BK_4})(t-s)]}{\epsilon_{\BK}+\epsilon_{\BK_2}-\epsilon_{\BK_3}-\epsilon_{\BK_4}}
    \\
    \times\,
    (
    N_{\BK\sigma}N_{\BK_2\bar{\sigma}}\bar{N}_{\BK_3\bar{\sigma}}\bar{N}_{\BK_4\bar{\sigma}}
    -
    \bar{N}_{\BK\sigma}\bar{N}_{\BK_2\bar{\sigma}}N_{\BK_3\bar{\sigma}}N_{\BK_4\sigma}
    )\Big|_s\,ds
    \,,
  \end{multline}
  \begin{multline}
    \label{eq:prethermeqhubb}
    N_{\BK\sigma}(t)
    =
    N_{\BK\sigma}(0)
    -
    4\frac{U^2}{L^2}
    \sum_{\BK_2\BK_3\BK_4}\Delta(\BK+\BK_2-\BK_3-\BK_4)
    \\\times
    \frac{\sin^2[(\epsilon_{\BK}+\epsilon_{\BK_2}-\epsilon_{\BK_3}-\epsilon_{\BK_4})t/2]}{(\epsilon_{\BK}+\epsilon_{\BK_2}-\epsilon_{\BK_3}-\epsilon_{\BK_4})^2}
    \\
    \times\,
    (
    N_{\BK\sigma}N_{\BK_2\bar{\sigma}}\bar{N}_{\BK_3\bar{\sigma}}\bar{N}_{\BK_4\bar{\sigma}}
    -
    \bar{N}_{\BK\sigma}\bar{N}_{\BK_2\bar{\sigma}}N_{\BK_3\bar{\sigma}}N_{\BK_4\sigma}
    )\Big|_0
    \,,
  \end{multline}
  \begin{multline}
    \label{eq:QBEhubb}
    \frac{dN_{\BK\sigma}(\tau)}{d\tau}
    =
    -\frac{2\pi}{L^2}
    \sum_{\BK_2\BK_3\BK_4}\Delta(\BK+\BK_2-\BK_3-\BK_4)
    \\\times
    \delta(\epsilon_{\BK}+\epsilon_{\BK_2}-\epsilon_{\BK_3}-\epsilon_{\BK_4})
    \\
    \times\,
    (
    N_{\BK\sigma}N_{\BK_2\bar{\sigma}}\bar{N}_{\BK_3\bar{\sigma}}\bar{N}_{\BK_4\bar{\sigma}}
    -
    \bar{N}_{\BK\sigma}\bar{N}_{\BK_2\bar{\sigma}}N_{\BK_3\bar{\sigma}}N_{\BK_4\sigma}
    )
    \Big|_\tau
    \,,
  \end{multline}
  where~\eqref{eq:prethermeqhubb} agrees with the prethermalization
  result in Ref.~\onlinecite{Moeckel2008a}, and $\tau$ $=$ $tU^2$.  
  It is straightforward to show that if one starts from an unpolarized
  initial state, $N_{\BK\uparrow}(0)$ $=$ $N_{\BK\downarrow}(0)$, no
  polarization is generated by the time evolution. In this case the
  spin indices can be omitted in the momentum distributions.

  In the limit of infinite lattice dimensions, the von-Laue function
  can be replaced by $\frac1L$~\cite{Muellerhartmann1989b}, and the
  each momentum summation can then be expressed as an energy integral
  over the density of states.  We only consider this case from now on,
  and furthermore assume particle-hole symmetry, i.e., a symmetric
  density of states $D(\epsilon)$ $=$ $D(-\epsilon)$, density $n$ $=$
  1, and Fermi energy $0$. As in the main text, the initial state is
  assumed to be the (unpolarized) ground state of $H_0$.

  \subsubsection{Effective temperature}

  As discussed in the main text, the effective temperature of the
  system is obtained by equating the kinetic energy in the final
  state, given to leading order in a Sommerfeld expansion by
  \begin{align}
    %E_{\text{kin}}^{\text{fin}}
    %=
    E_{\text{kin}}(\tau=\infty)
    &=
    2\int D(\epsilon)n_F(\epsilon)\,\epsilon\,d\epsilon
    \\
    &=
    E_{\text{kin}}^{\text{ini}}+\frac{\pi^2}{3}D(0)T^2+O(T^4)
    \,,
    \intertext{with the kinetic energy of the prethermalized state, which can be written as~\cite{Eckstein2010a}}
    %E_{\text{kin}}^{\text{pre}}
    %=
    E_{\text{kin}}(\tau=0^+)
    &=
    E_{\text{kin}}^{\text{ini}}+2S_4U^2+O(U^4)
    \,,
    \intertext{with the abbreviation}
    S_4
    &=
    \int\limits_0^\infty\IM\left[\int\limits_0^\infty D(\epsilon)e^{iu\epsilon}d\epsilon\right]^4du
    \,,
  \end{align}
  so that the kinetic energy of the initial state
  $E_{\text{kin}}^{\text{ini}}$ cancels. All kinetic energies are given
  per lattice site.
  Solving for $T$ gives
  \begin{align}
    T&=aU+O(U^2)
    \,,
    &
    a&=\sqrt{\frac{6S_4}{\pi^2D(0)}}
    \,.
  \end{align}
  For the semielliptical density of states with bandwidth 4, we have
  $D(0)$ $=$ $1/\pi$, $S_4$ $=$ $0.0209$, $a$ $=$ $0.200$.

  \subsubsection{Relaxation rate}
  
  Replacing all momentum occupation numbers by their thermal Fermi
  function values, as described in the main text,
  reduces~\eqref{eq:QBEhubb} near the Fermi energy to 
    \begin{align}
    \frac{\delta n^\pm}{d\tau}
    &=-\widetilde{\gamma}_{\text{F}}\,\delta n^\pm(\tau)+O(\delta n^\pm(\tau)^2)
    \,,
  \end{align}
  with $\widetilde{\gamma}_{\text{F}}$ given in terms of the
  phase-space integral
  \begin{align}
    \frac{\widetilde{\gamma}_{\text{F}}}{4\pi}&=
    \int\limits_{-\infty}^{\infty}
    \int\limits_{-\infty}^{\infty}
    \int\limits_{-\infty}^{\infty}
    D(\epsilon_2)
    D(\epsilon_3)
    D(\epsilon_4)\,
    \delta(\epsilon_2-\epsilon_3-\epsilon_4)
    \nonumber\\&~~~~\times
    (1-n_{\text{F}}(\epsilon_2))
    n_{\text{F}}(\epsilon_3)
    n_{\text{F}}(\epsilon_4)\,
    d\epsilon_2
    d\epsilon_3
    d\epsilon_4
    \nonumber\\
    &=
    \int\limits_{-\infty}^{\infty}
    \left[
      \int\limits_{-\infty}^{\infty}
      \frac{\cos(\lambda\epsilon)}{2\cosh(\beta\epsilon/2)}
      D(\epsilon)
      d\epsilon
    \right]^3
    \frac{d\lambda}{2\pi}
    \nonumber\\&
    =
    T^2
    \int\limits_{0}^{\infty}
    \left[
      \int\limits_{0}^{\infty}
      \frac{\cos(ux)}{\cosh(x/2)}
      D(Tx)
      dx
    \right]^3
    \frac{du}{\pi}
    \nonumber\\&
    =
    \frac{\pi^2}{4}D(0)^3T^2+O(T^4)
    \,,
  \end{align}
  which is proportional to $T^2$ because of the softening of the Fermi
  functions for $T$ $>$ 0.  Here we used the delta function
  representation $\delta(x)$ $=$ $\int_{-\infty}^{\infty}d\lambda
  e^{i\lambda x}/(2\pi)$ and the symmetry of the density of states,
  $D(\epsilon)$ $=$ $D(-\epsilon)$. In the integrals we substituted
  $x$ $=$ $\beta\epsilon$ and $u$ $=$ $\lambda T$. The last square
  bracket yields $\pi D(0)\text{sech}(\pi u)$ $+$ $O(T^2)$. Then
  finally $\int_{0}^{\infty}\text{sech}(x)^3dx$ $=$ $\pi/4$. Scaling
  back from $\tau$ to $t$ yields $\gamma_{\text{F}}$ $=$
  $\widetilde{\gamma}_{\text{F}}U^2$ as given
  in the main text.

\end{document}